\documentclass[3p,times,procedia]{elsarticle}
\usepackage{nupha_ecrc}

\volume{00}

\firstpage{1}

\journalname{Nuclear Physics A}

\runauth{}

\jid{nupha}

\jnltitlelogo{Nuclear Physics A}

\usepackage{amssymb}

\usepackage[figuresright]{rotating}

\begin{document}

\begin{frontmatter}



\dochead{XXVIIth International Conference on Ultrarelativistic Nucleus-Nucleus Collisions\\ (Quark Matter 2018)}

\title{Precision Dijet Acoplanarity Tomography of the
Chromo Structure of Perfect QCD Fluids}


\author[LBL,Wigner,CU,CCNU]{M. Gyulassy}
\author[Wigner]{P. Levai}
\author[IU,CCNU]{J. Liao}
\author[IU]{S. Shi}
\author[LBL]{F. Yuan}
\author[LBL,CCNU]{X.N. Wang}

\address[LBL]{Nuclear Science Division, Lawrence Berkeley National Laboratory, Berkeley, CA 94720, USA}

\address[Wigner]{MTA 
Wigner Research Center for Physics, 1525 Budapest, 
  Hungary}

\address[CU]{Pupin Lab MS-5202, Department of Physics, Columbia University, New York, NY 10027, USA}

\address[CCNU]{Institute of Particle Physics, Central China Normal University, Wuhan, China}


\address[IU]{Physics Department and Center for Exploration of Energy and Matter,
Indiana University, \\ 2401 N Milo B. Sampson Lane, Bloomington, IN 47408, USA}

\begin{abstract}Dijet acoplanarity is dominated by vacuum (Sudakov) pQCD radiation even in Pb+Pb collisions, but future higher precision measurements of the tails of the acoplanarity distributions can help to resolve {\em separately} the medium opacity, $\chi=L/\lambda$, and the color screening scale $\mu^2$ from the path averaged BDMS saturation scale $Q_s^2[\chi,\mu]=\int dL\; \hat{q}(E,T(L))\propto
  \mu^2 L/\lambda $, that is already well constrained by 
  nuclear modification factor data on $R_{AA}(p_T)$. We compare Gaussian (BDMS) and non-Gaussian (GLV) models of medium broadening of vacuum (Sudakov) induced acoplanarity distributions in A+A.   With few percent accuracy on the ratio of A+A to p+p distributions, experiments can easily identify non-Gaussian Landau and Rutherford tails due to multiple collisions. However, we find that sub-percent precision will be
  required to constrain $\chi$ and $\mu$ separately from $Q_s$.
  
\end{abstract}


\end{frontmatter}


\section{Introduction}
\label{}

Nearly back-to-back di-jets with transverse momenta
$(\vec{P}_1=\vec{Q}+\vec{q}/2\;\vec{P}_2=-\vec{Q}+\vec{q}/2)$, become acoplanar even in the vacuum due to multi-gluon radiation. Here $\vec{q}=\vec{P}_1+\vec{P}_2$ is the net dijet acoplanarity transverse momentum. We consider the kinematic range where $P_1\approx P_2 \approx Q \gg Q_s =\sqrt{\langle \chi \mu^2 \rangle}$. Consider a jet with color Casimir and flavor $a$ traversing  a multi-component dynamic medium with evolving densities
 $\rho_b(x,t)$ with $b=q,g,m$.  The medium opacity $\chi$, the effective chromo screening scale, $\mu^2$, the  medium saturation momentum scale, $Q_s$, and the evolving jet transport coefficient fields, $\hat{q}_a(x,t)$ 
are related via
\begin{eqnarray}
  \chi &=&L/\lambda_a\equiv \langle  \int dt \sum_b \rho_b (z(t),t) \int dq_\perp d\sigma^{ab}/dq_\perp^2\rangle \\
  Q_s^2&=&\chi\mu^2\equiv\langle\int dt\; \hat{q}_a(z(t),t)\rangle
  \equiv \langle \int dt \sum_b\rho_b(z(t),t) \int dq_\perp^2 \; q_\perp^2 d\sigma^{ab}/dq_\perp^2\rangle \;\;.
\end{eqnarray}
where $\langle\cdots\rangle$ denotes an ensemble average over jet
paths $z(t)$ and over evolving medium chromo electric and magnetic
quasi-parton densities in a given experimentally defined centrality class,
${\cal C}=\{dN_{ch}/d\eta,v_n^{soft},\cdots\}$.  These functionals depend on
the viscous hydrodynamic evolution of the component densities
as well as on the microscopic details of jet
medium interactions modeled here by the multi-channel
$d\sigma^{ab}$ differential cross section of the jet parton $a$ with
quasi-partons of type $b$ in the medium. The goal of measuring correlations
between hard $p_T>10$ GeV jet observables  and soft $p_T<2$ GeV
observables is to experimentally constrain and discriminate between
different models of the chromo-structure of the near perfect
QCD fluids produced in
ultra relativistic nuclear collisions at RHIC and LHC. Recent
efforts to discriminate between weakly coupled wQGP and strongly coupled sQGMP (semi-Quark-Gluon-Monopole-Plasma)
models of the chromo structure of QCD fluids 
were discussed in refs.\cite{Jaki,CIBJET,Xu:2014tda}. 
The most discriminating {\em Soft$\otimes$Hard} correlation observable
measured so far appears to be the $\sqrt{s}=0.2,2.76,5.02$ ATeV and centrality class {$\cal C$}
dependence of the jet-medium elliptic asymmetry
$v_2^h(p_T)\equiv \langle
v_2^{soft}v_2^{hard}(p_T>10)\rangle/\langle (v_2^{soft})^2\rangle$.
Many models are already ruled out by demanding a consistent simultaneous
account of both hard jet observables as well as soft flow azimuthal asymmetry observables.

However, even state of the art event-by-event soft-hard frameworks
such as ebe-vUSPH-BBMG\cite{Jaki} and ebe-VISNU+CUJET3\cite{CIBJET,Xu:2014tda},
that can account consistently and simultaneously for all the current combined soft+hard data $\{RHIC+LHC\}\otimes\{R_{AA}^{soft},R_{AA}^{hard},v_n^{soft}(p_T<2),v_n^{hard}(p_T>10)\}\otimes\{u,d,s,c,b\}$ at the $(\chi^2/dof)<2$ level,
have not yet been able to converged on the fundamental physics
questions related to
the  relevant chromo electric and magnetic
degrees of freedom $\rho_b$ and the microscopic details of $d\sigma^{ab}$
needed to explain simultaneously
 the observed near perfect fluidity of the bulk 
 and the correlated hard jet and dijet azimuthal asymmetries.

 These open
questions motivate the present exploratory
study to test whether dijet acoplanarity observables could help to ``illuminate'' the chromo structure of perfect QCD fluids.
In particular, can acoplanarity distribution shape analysis
be used 
(in addition to extensive data on observables listed above, e.g.,see fig 1 of \cite{CIBJET})
to determine experimentally the opacity $\chi$ separately from
the rms mean transverse momentum scale $Q_s$ \cite{jacobs} ?  As shown in
\cite{CIBJET,Xu:2014tda} the emergent color magnetic monopole
component near $T_c$ as constrained by lattice QCD data  leads to critical
opalescence like enhancement of $\hat{q}(T,E)$ near $T_c$ 
as originally proposed in ref.\cite{liao}. This arises from the Dirac
constraint on electric and magnetic monopole couplings,
$\alpha_E\alpha_M=1$, that leads to $d\sigma^{qm} \sim
1/\alpha_E^2 \; d\sigma^{qg} \gg d\sigma^{qg}$~\cite{Liao:2006ry}.  In
\cite{CIBJET,Xu:2014tda} it was shown that with model parameters
constrained by lattice QCD data as well as global A+A $(\chi^2/dof)$ data analysis,
the jet transport field ratio $\hat{q}(T,E>10)/T^3$ not only
maximizes near $T_c$ but also 
the viscosity to entropy ratio $\eta/s\sim
T^3/\hat{q}(T,3T)\approx 0.1-0.2$,  minimizes near $T_c$ close to the quantum
(holographic KSS) bound~\cite{Danielewicz:1984ww,Kovtun:2004de}. The sQGMP chromo structure is therefore not
only $(\chi^2/dof)<2$ consistent with available data including the jet $v_2^h(p_T)$, but it also naturally accounts
for the near perfect fluidity property of QCD  matter produced at
RHIC and LHC. So, can acoplanarity help to support or to falsify this picture? 

Existing data from  RHIC\cite{phenixstar} and LHC\cite{ALICE15}
on dijet acoplanarity are encouraging but suffer from too large uncertainties to be useful. Fortunately new jet finding techniques
are being developed\cite{jacobs,STAR17}
to reduce background fluctuations and to achieve
higher precision measurements.
Jet-medium corrections in A+A to the dominant
vacuum (p+p) acoplanarity  decrease rapidly with increasing
dijet momenta $Q$. Our main conclusion below is that sub percent precision on p+p as well as A+A
will be  required to resolve separately the opacity $\chi=L/\lambda$ and the effective screening scale $\mu$ from $Q_s$.
The most favorable  kinematic range appears to be the moderate
$10< Q<30$ GeV window above the bulk collective flow and intermediate hadron
recombination range
but not too high to enable extraction of the small jet-medium broadening
signal ($\propto \chi\mu^2/Q^2$, see  Fig.1 below).

The  dijet relative azimuthal acoplanarity angle, $\phi=\phi_1-\phi_2$, is approximately linearly related to the dijet acoplanarity transverse momentum $q(Q,\Delta\phi) \approx Q\;(\pi- \phi)$ in the  angular range $\{3\pi/4,\pi\}$.
We calculate the azimuthal acoplanarity distribution convoluting perturbative QCD  gluon showers with medium induced broadening.
We utilize the impact parameter, $b$, representation 
\cite{Appel,Baier:1996sk,Eramo13,Luo:2018pto,Mueller:2016,Chen:2016vem}. 

In this exploratory 
study, we apply  the formalism  developed
by Mueller et al \cite{Mueller:2016} and Chen et al \cite{Chen:2016vem}
with the main difference that instead of treating
jet medium induced acoplanarity broadening
in the BDMS (Gaussian) approximation\cite{Baier:1996sk} (see eq.5 of\cite{Chen:2016vem}),  we  calculate acoplanarity broadening
utilizing the non-Gaussian GLV elastic multiple
collision series (eqs.(21,23) of\cite{Gyulassy:2002yv}). Our aim is to compute the effect of
non-Gaussian intermediate Landau as well as hard  Rutherford tails contributions on the convoluted acoplanarity distribution shape at finite opacity $\chi=L/\lambda\sim 10$ and $\mu\sim 0.$ GeV.
As in \cite{Mueller:2016, Chen:2016vem},
the convolution of vacuum and medium sources of acoplanarity distributions
is computed via
\begin{equation}
   \frac{dN}{dq^2} \approx \frac{1}{Q^2}
  \frac{dN}{d \Delta \phi}=\int b db J_0(|q(Q,\Delta \phi)| b) e^{-S_{vac}(Q,b)-S_{med}(Q,b)} \;\;\;\; .
\end{equation}
We use eq.3 and parameters of \cite{Mueller:2016} to evaluate
the vacuum Sudakov factor 
\begin{equation}
  S_{vac}\approx (\alpha/2\pi) \sum_{q,g} \left\{(A_1 (\log(Q^2/\mu_b^2)^2/2 +(B_1 +D_1\log(1/R^2))\log(Q^2/\mu_b^2)\right\}+S_{NP}(Q,b)\;\;\;\; .
\end{equation} 
and their phenomenological nonperturbative $S_{NP}$ factor.  For illustration,
we take the jet radius to be $R=0.4$ and set vacuum Sudakov $\alpha\approx 0.09$ that fits approximately
the shape and normalization of azimuthal acoplanarity distribution observed in RHIC $p+p$ reactions.

Below we compare  medium induced broadening  assuming the one parameter BDMS\cite{Mueller:2016} $S_{BDMS}(b;Q_s)=b^2 Q_s^2/4$ approximation to that assuming
the two parameter GLV\cite{Gyulassy:2002yv}  $S_{GLV}(b;\chi,\mu)=\chi(\mu b K_1(\mu b)-1)$ multiple collision approximation.
The  GLV form used below assumes for simplicity that
all $ab$ channel have identical Yukawa screening scales, $\mu$. More generally,
in the CIBJET=ebe-CUJET3  sQGMP color composition framework\cite{CIBJET,Xu:2014tda} 
we  would use lattice QCD data to fix
the $T$ dependence of q,g,m densities and different
chromo electric and magnetic screening masses. 

In Fig.1a left panel numerical results with BDMS and GLV in
the absence of vacuum Sudakov acoplanarity are compared. The most obvious
point is that Yukawa multiple collision distributions have a power
law high $q\gg Q_s(\chi,\mu)$ ``Rutherford tail'', $dN/dq^2\sim \chi\mu^2/q^4$.
That Rutherford tail that extends far beyond the Gaussian BDMS
approximation. Importantly, for intermediate $q\sim Q_s$, the GLV
distribution is concave relative to the convex shape of the BDMS
Gaussian approximation.  This finite medium size 
$\chi=L/\lambda$ intermediate ``Landau tail''  of course gradually evolves
into the higher $q\gg Q_s$ Rutherford tail. However, as also shown in Fig.1a,
after convoluting $S_{vac}$
with $S_{med}$, the final acoplanarity distribution is always dominated by the
vacuum Sudakov tail  $\sim \alpha/q^2$.
The QCD vacuum conformal tail dominates acoplanarity distribution tails
even for moderate $Q=20$ GeV jets because  $Q_s^2=\langle
\hat{q}L\rangle =\langle \chi \mu^2\log(Q^2/\mu^2) \rangle \sim 10$ GeV$^2
\ll Q^2$ for realistic nuclei.

In Fig 1b the dijet azimuthal acoplanarity distribution $dN/d\Delta\phi$
is compared for two illustrative sets of parameters $Q_s^2=10$ and $16$ GeV$^2$  for vac+BDMS in blue (dash vs solid) and for vac+GLV for fixed $\mu=0.5$ GeV
and $\chi=6$ vs $10$ in red (dash vs solid). The black curve is the vacuum Sudakov distribution in this case. Note that with these parameters the $\Delta \phi=\pi$ intercept is identical for BDMS$(Q_s^2=9.6)$ and GLV$(\mu=0.5,\chi=10,Q_s^2=16)$, but the curves intertwine slightly as $\Delta\phi$ varies away from back-to-back $\Delta\phi=\pi$ intercept point. 
It is immediately clear that very high precision will be needed experimentally
to discriminate between BDMS versus GLV medium broadening once the
intercept at $\pi$ is constrained.

To get a more quantitative feeling about the sensitivity level
required to resolve
$\chi$ and $\mu$ from the distribution shapes, we show in Fig. 2a the (vac+med)/vac ratio, $R(q)$. On left panel BDMS and GLV $Q_s$ are separately adjusted
as in Fig.1b to fit the vacuum acoplanarity intercept $R(0)\approx 0.6$. In this plot the shape of vac+med {\em relative} to the vacuum distribution shows
three characteristic feature due to  medium broadening: (1) $R(0)<1$, (2) there is a point $q=q_{med}$ where $R(q_{med})=1$, and (3) there is a point $q\equiv q_{Ruth}$ where  $R_{BDMS}(q_{Ruth})=R_{GLV}(q_{Ruth})$
and the relative ordering of BDMS and GLV switches sign to  $1< R_{BDMS}(q)<R_{GLV}(q)$ above $q_{Ruth}$. As expected, GLV has the longer Rutherford enhancement of the vacuum acoplanarity.

In Fig. 2b, the sensitivity of vac+GLV acoplanarity to variations of
$\mu=0.25,0.50,0.75$ with $Q_s=9.6$ or $16$ GeV$^2$ is shown. As can be
seen, very high sub-percent precision level would be needed of the ratio of
A+A to p+p acoplanarity distributions  
to be able to resolve $(\chi,\mu)$ from the shape
analysis of dijet distributions. In contrast, from Fig.2a,
it is clear that only a few percent precision
could suffice distinguish between Gaussian
and GLV multiple collision shapes.
That level of high precision is similar to discriminate
between ebe models of soft-hard $v_3(p_T>10)$ correlations\cite{CIBJET}.
We note finally
that for 
acoplanarity angles larger than considered here, addition important dynamical contribution from multiple jet-medium secondary interactions must be taken into account (see esp. Fig.6 of \cite{Luo:2018pto}).

\begin{figure}[!hbt]
\begin{center} \vspace{-0.1in}
\includegraphics[width=0.42\textwidth]{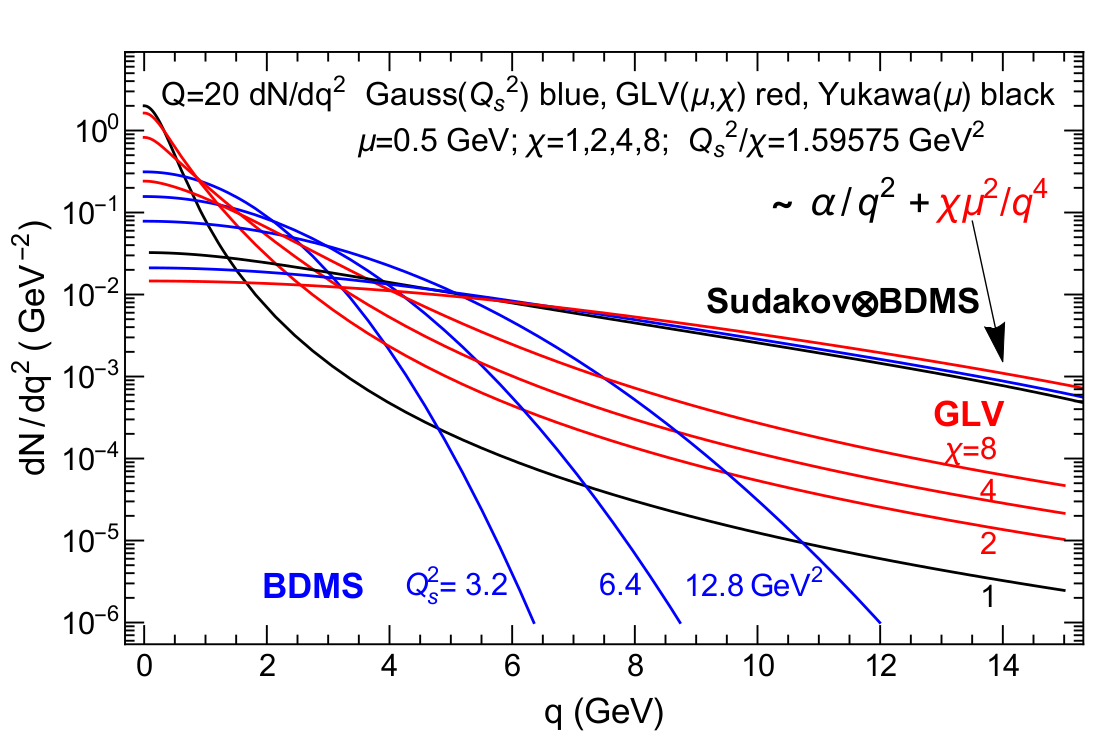}
\includegraphics[width=0.42\textwidth]{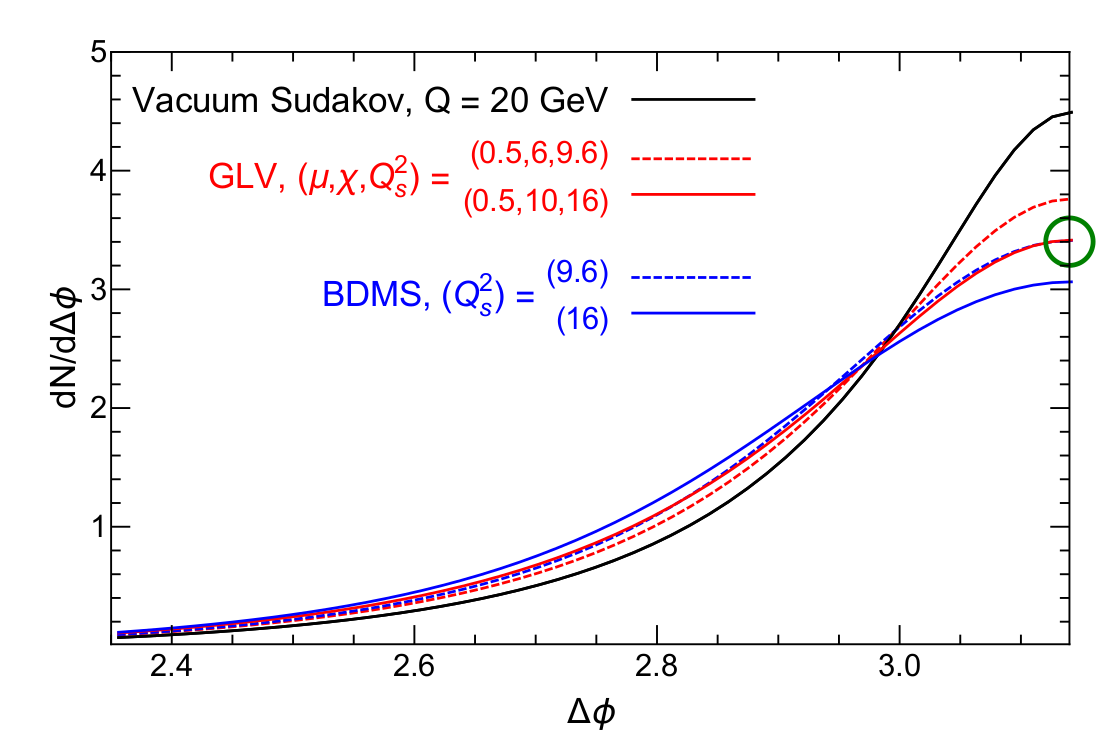}
\vspace{-0.15in}
\caption{(color online) (a) Jet-Medium multiple scattering broadening of dijet $dN/dq^2$ in BDMS(blue) and GLV(red) approximations without vacuum effects
  compared to convoluted vacuum+medium distribution for
  $Q_s^2=3.2,6.4,12.3$ GeV$^2$. (b) The dijet azimuthal $dN/d\Delta \phi$ distributions for $Q_s^2=9.6$ (dash) and $16$ (solid) are compared to vacuum p+p (black) for BDMS (blue) and GLV (red). Note that with the intercept at $\Delta\phi=\pi$ constrained (circled point), very high precision is needed to differentiate
  BDMS and GLV.}  \label{fg1}

\end{center}
\end{figure} 

\begin{figure}[thb]
\begin{center} \vspace{-0.25in}
 \includegraphics[width=0.42\textwidth]{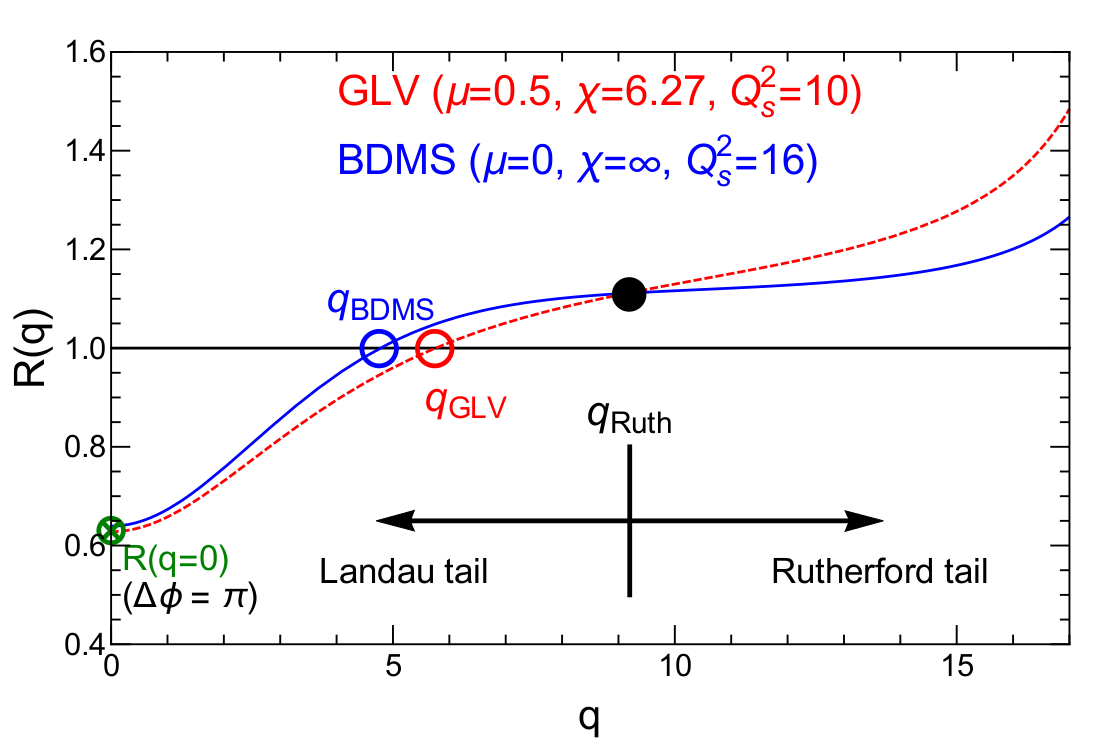}
\includegraphics[width=0.42\textwidth]{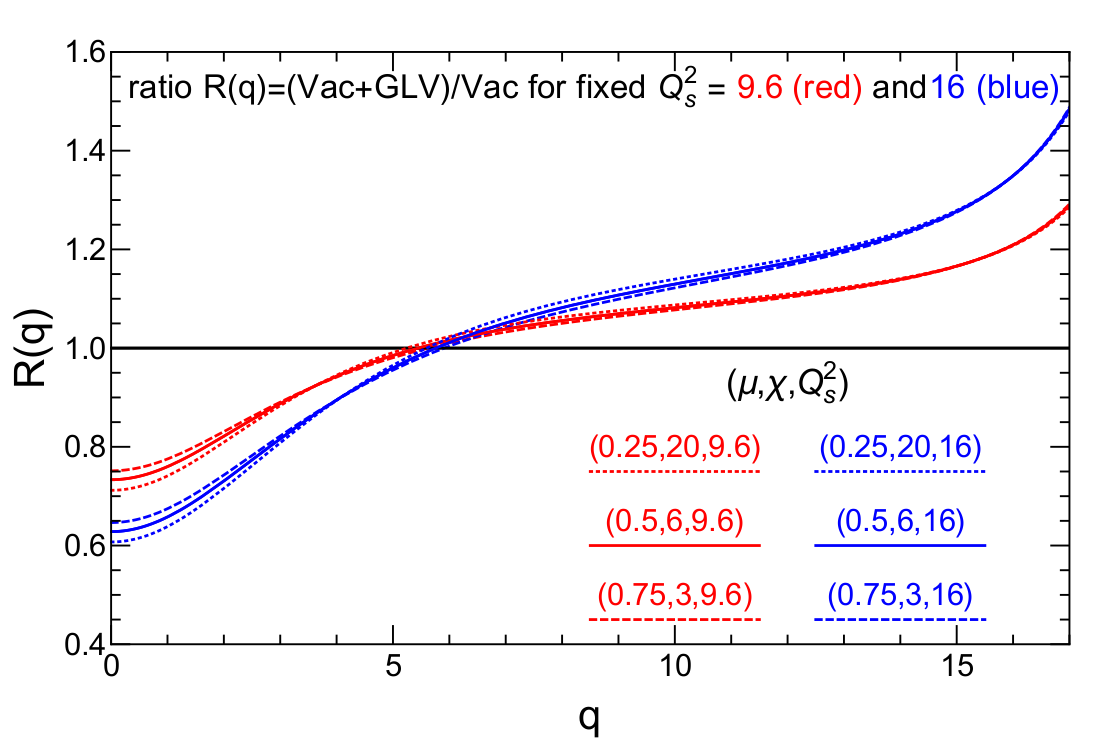}
\vspace{-0.15in}
\caption{(color online) (a) Ratio $R(q)$ of vacuum+medium $dN/dq^2$ to vacuum
  acoplanarity dijet momentum distributions for $Q=20$ GeV and $\alpha_s=0.09$. Blue curve
  illustrates acoplanarity  distribution shape ratio with BDMS($Q_s^2=16$) 
  while  Red is with GLV$(\mu=0.5,\chi=6.27)$ with $\chi$
  adjusted to coincide with the BDMS intercept at $R(q=0)\approx 0.6$.
  The BDMS approximation is broader than the finite opacity GLV$(\mu,\chi)$ due to the  concave ``Landau tail'' of GLV (see Fig.1a)  
  up to $q<q_{Ruth}\approx 10$ GeV. But GLV is broader than BDMS in the Rutherford tail region
  $q>q_{Ruth}$. (b) Shows the magnitude of variation (Vac+GLV)/Vac
  to variations of $\mu=0.25,0.50,0.75$ GeV for $Q_s^2\approx 10$ (red) and $16$ (blue) GeV$^2$. Sub-percent level precision would be required to  resolve $\chi$ and $\mu $ from $Q_s^2 \approx \chi\mu^2 \log(Q^2/\mu^2)$.}
  
\label{fg2}

\end{center}
\vspace{-0.2in}
\end{figure}


{\small {\bf Acknowledgments.} 
Special thanks to Peter Jacobs for 
discussions about precision dijet acoplanarity observables. The research of JL and SS is supported by  the NSF Grant No. PHY-1352368. FY and XNW are supported by  DOE grant DE-AC02-05CH11231, and PL is supported in part by OTKA grant K120660. MG, JL, and XN are partially
 supported by the IOPP, CCNU, Wuhan, China  NSFC grants
11775095, 11375072, 1122150 and 11735007 and MTA CH-HU.} 
\vspace{-0.2in}




\bibliographystyle{elsarticle-num}
\bibliography{<your-bib-database>}



\end{document}